\documentclass[twocolumn]{aastex631}
\usepackage{multirow}

\newcommand{\ls}{LS~I~+~61$^\circ$303}
\newcommand{\gr}{$\gamma$-ray}

\shortauthors{Zhang}
\graphicspath{{./}{figures/}}

\begin{document}

\title{A New Puzzling Periodic Signal in GeV Energies of the $\gamma$-Ray Binary LS I +61$^\circ$303}

\author{Pengfei Zhang}
\affiliation{Department of Astronomy, School of Physics and Astronomy, Key Laboratory of Astroparticle Physics of Yunnan Province, Yunnan University, Kunming 650091, People's Republic of China; zhangpengfei@ynu.edu.cn}

\begin{abstract}
\ls~is a high-mass X-ray binary system comprising a massive Be star and a rapidly rotating neutron star.
Its spectral energy distribution across multi-wavelengths categorizes it as a \gr~binary system.
In our analysis of \ls~using \emph{Fermi}-LAT observations, we not only confirmed
the three previously discussed periodicities of orbital, superorbital, and orbital-superorbital
beat periods observed in multi-wavelength observations, but also identified an additional periodic signal.
This newly discovered signal exhibits a period of $\sim$26.3~day at a $\sim7\sigma$ confidence level.
Moreover, the power spectrum peak of the new signal gradually decreases as the energy increases
across the energy ranges of 0.1–0.3, 0.3–1.0, and 1.0–500.0~GeV.
Interestingly, a potential signal with a similar period was found in data obtained from
the Owens Valley Radio Observatory 40~m telescope.
We suggest that the newly discovered periodic signal may originate from a coupling between
the orbital period and the retrograde stellar precession period.
\end{abstract}


\section{Introduction}
\label{Intro}
\ls~is a high-mass X-ray binary system, composed of a young massive Be star \citep{gcg+07}
and a rapidly rotating neutron star \citep{wen+22},
with a non-thermal electromagnetic emission extending from MHz radio frequencies to TeV \gr~energies \citep{d13}.
Its properties of the spectral energy distribution (SED) in multi-wavelength emissions indicate
that \ls~ is dominated by the MeV-GeV \gr s, this makes it to be a \gr~binary.
Until now, very few \gr~binary systems have been found to produce detectable \gr~emissions,
with only a handful in our Galaxy \citep{aha+05a,aha+05b,alb+06,hin+09,lat+12,cor+19}
and one in the Large Magellanic Cloud \citep{cor+16}.

\ls~has a orbital period of $P_1\sim26.496$~day \citep{g02} with an eccentricity
$e\sim0.54$ \citep{ara+09} and a companion mass of 12.5~$M_{\Sun}$ \citep{cas+05}, and locates at a
distance of 2.0 kpc \citep{fh91}.
The zero point of its orbital phase ($\phi$) has historically been defined at JD~=~2,443,366.775
(i.e. $\phi_0$) in \citet{g02}, and the orbital phase of its periastron position is $\phi\sim0.275$,
which adopted from \citet{ara+09}.
In addition to the orbital period of \ls, a long-term modulation period of
1667 day ($P_2$) was discovered in GHz radio observations by \citet{g02}.
With the increasing accumulation of observational data for \ls, \citet{mj13} reported a third
modulation period of 26.92 day ($P_3$) in 6.7 years Green Bank Interferometer radio database at
2.2 GHz and 8.3 GHz,
and this period was also revealed in 0.1--300.0~GeV with the \emph{Fermi}-LAT observations
by \citet{jm14,cmn+23}.
Regarding the potential origin of the final two periods, previous literature offered extensive
discussions \citep{g02,mj13,jm13,mt14,jm14,mjh15,mt16}.
Based on \emph{Fermi}-LAT GeV observations for \ls, \citet{xwt17} suggested that a non-axisymmetric
circumstellar disk may be present around the Be companion, which rotates with a period of 1667 days,
leading to the long-term modulation,
and the period of 26.92 days is a result of the beat frequency ($f_3=\frac{1}{f_1-f_2}$)
between the orbital ($P_1$) and long-term ($P_2$) periods.

In GeV, \ls~has a \gr~counterpart named J0240.5+6116 in the first \emph{Fermi}-LAT source catalog
\citep[1FGL;][]{1fgl2010} and J0240.5+6113 in the fourth catalog
Data Release 4 \citep[4FGL-DR4;][]{4fgl-dr3,4fgl-dr4}.
Based on it, we analyzed the \gr~events around \ls~from
the $\sim$15~year \emph{Fermi}-LAT observations. In our timing analysis, besides three periodicities
that have been reported in previous literature, an additional new periodic signal reveals at
$26.301\pm0.037$~day at a $\sim7\sigma$ confidence level. Interestingly, a potential signal
has a similiar period at $\sim26.16\pm0.11$~day shown in the Owens Valley Radio Observatory (OVRO)
data in the Figure~3~(d) of \citet{jar+18}, but it is not significant.
The following presents our data analysis and results.

\section{Data Analysis and Results}
\label{sec:lat-data}

\subsection{Data Reduction}
\label{sec:model}

\begin{table*}
\begin{center}
\caption{Best-fit results of likelihood analysis}
\begin{tabular}{ccccccc}
\hline\hline
Models & \multicolumn{6}{c}{Parameter values}  \\
\hline
LP & $\alpha$ & $\beta$ & $E_{\rm b}$ & TS & $F_{\rm ph}$ & $F_{\rm en}$ \\
   & 2.461$\pm$0.006     & 0.112$\pm$0.002     & 1.517 & 227111.0 & $7.861\pm0.030$ & 4.535$\pm$0.014 \\
   & 2.445$\pm0.015^\phi$ & 0.117$\pm0.006^\phi$&1.517$^\phi$ & 31694.1$^\phi$ & 7.283$\pm$0.083$^\phi$ & 4.308$\pm$0.047$^\phi$\\
\hline
PLEC & $\Gamma$ & $b$ & $E_{\rm c}$ & TS & $F_{\rm ph}$ & $F_{\rm en}$ \\
     & 1.979$\pm$0.015 & 0.605$\pm$0.026 & 2.927$\pm$0.351 & 226614.0 & 7.793$\pm$0.032 & 4.498$\pm$0.017\\
\hline
\end{tabular}
\label{tab:par}
\end{center}
{\bf Notes. }{Best-fit parameter values of the likelihood analysis in 0.1--500.0~GeV
             for LP and PLEC models, $^\phi$values derived from the dip data, $E_{\rm b}$ and
             $E_{\rm c}$ in units of GeV. $F_{\rm ph}$ is the integrated photon
             flux in units of $10^{-7}$~photons~cm$^{-2}$~s$^{-1}$ and $F_{\rm en}$
             is the integrated energy flux in units of $10^{-10}$~erg~cm$^{-2}$~s$^{-1}$.}
\end{table*}

We carried out a whole data analysis by
selected the \emph{Fermi}-LAT Pass 8 \emph{Front+Back} events (evclass = 128 and evtype = 3)
in the energy range of 0.1--500.0~GeV within a $20^\circ\times20^\circ$ region of interest (RoI)
centered at 4FGL~J0240.5+6113
(R.~A.~=~$\rm02^h40^m34^s.22$ and decl.~=~$+61^{\circ}13'43\farcs30$).
The data observations span from 2008 August 4 to 2023 September 21 (MJD: 54682.687--60208.046).
We removed the events with zenith angles $>90^\circ$ to exclude the \gr~contamination from the
Earth Limb, those with quality flags of “bad”, by a expression of
“DATA\_QUAL$>$0~\&\&~LAT$\_$CONFIG==1” to save high-quality events in good time intervals.
The instrumental response function “P8R3\_SOURCE\_V3” and
the latest available Fermitools-2.2.0 were used in the following analysis.
\begin{figure}
\centering
\includegraphics[angle=0,scale=0.56]{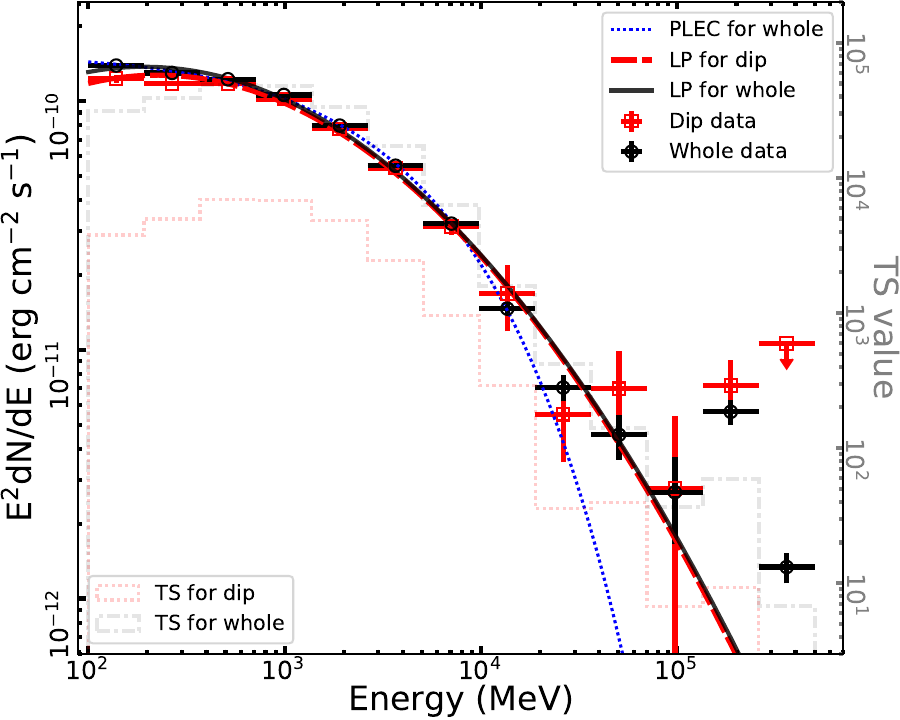}
\caption{\gr~SEDs of \ls~in 0.1--500.0~GeV. The best-fit LP models of the dip
         and whole LAT data are shown as red dashed and black solid lines, respectively.
         The PLEC model is shown with a blue dotted line.
         The gray dashed-dotted and red dotted histograms stand for TS values of
         data from the dip and whole LAT.}
\label{fig:sed}
\end{figure}
\begin{figure*}
\centering
\includegraphics[angle=0,scale=0.66]{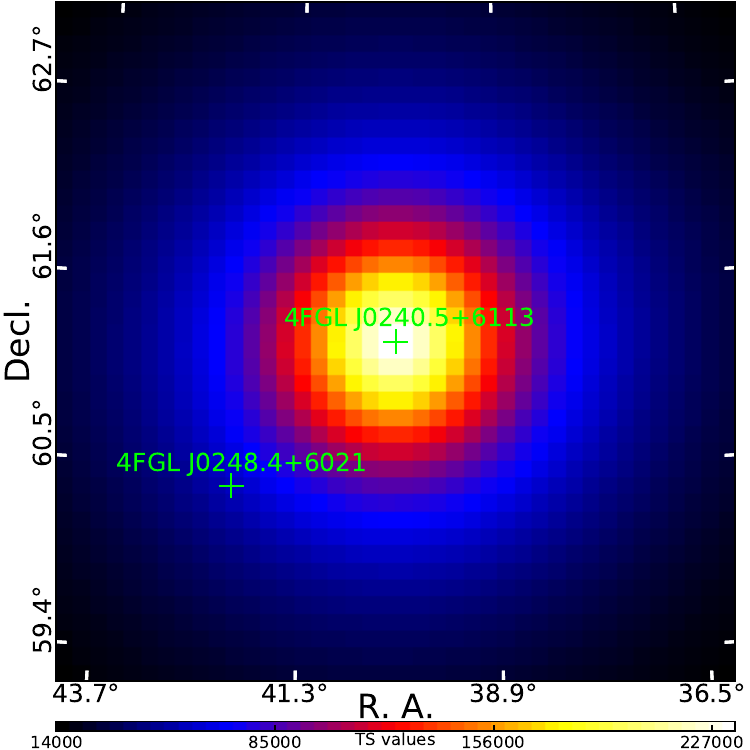}
\includegraphics[angle=0,scale=0.66]{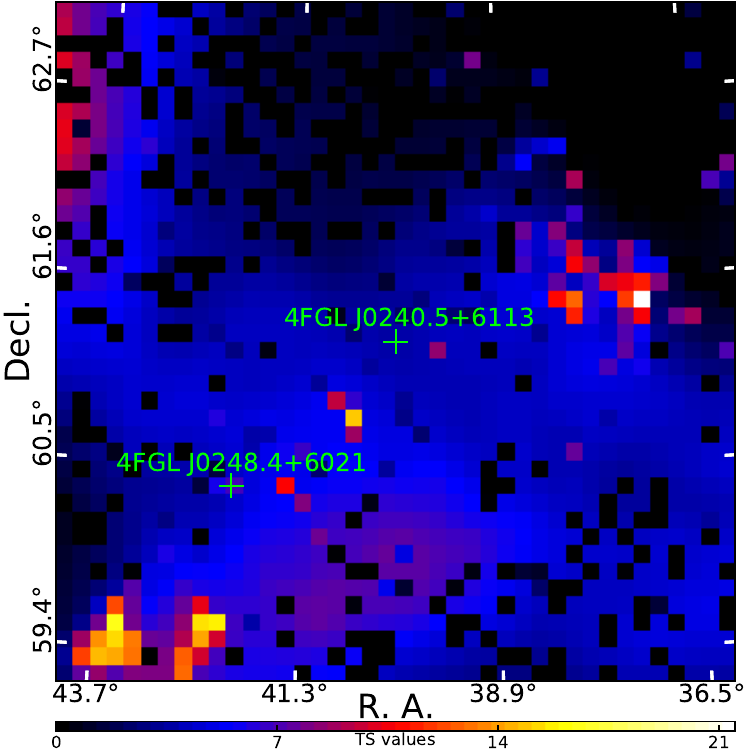}
\caption{TS maps in 0.1--500.0~GeV covering a $4^{\circ}\times4^{\circ}$ region centered at
         4FGL~J0240.5+6113 with a pixel of $0^{\circ}.1$. The \gr~sources reported in the 4FGL-DR4
         are shown with the green crosses.
         Left panel: TS map standing for the \gr~emissions from \ls~was created by fixing all model
         parameters in the new model and removing 4FGL~J0240.5+6113 from the model.
         Right panel: residual TS map was created based on the same model with target having
         a LP spectral shape,  with the exception that 4FGL J0240.5+6113 is retained.}
\label{fig:tsmap}
\end{figure*}

A model file was created by a python script make4FGLxml.py based on the newest 4FGL-DR4 catalog,
which includes all the parameters of the sources within $25^\circ$ centered at 4FGL~J0240.5+6113.
Then we modified the file to free the flux normalizations and spectral parameters for
the sources within $5^\circ$, the normalizations for the sources within $5^\circ$--$10^\circ$,
together with the ones outside $10^\circ$ but identified as variable sources.
The two normalizations for the diffuse emission components of Galactic and extragalactic were also set free.
All other parameters were fixed to be their values as them provided in 4FGL-DR4.
Then a binned maximum likelihood analysis was performed to update the free parameters
by employed the 15~yr \emph{Fermi}-LAT data. The best-fit parameters were saved as a new model file,
the following analysis based on this model. For \ls, a spectral shape of a log-parabola model (LP),
$dN/dE=N_0(E/E_b)^{-[\alpha+\beta\log(E/E_b)]}$, is provided in 4FGL-DR4.
The best-fit parameters of \ls~are summarized in Table~\ref{tab:par}.
Considering that a pulsar hosted in \ls, we also used typical pulsar \gr~model of a power-law with
an exponential cutoff (PLEC), $dN/dE=N_0(E/E_0)^{-\Gamma}\exp[-(E/E_{\rm c})^b]$,
to describe the target's \gr~emission.
The best-fit parameters of the PLEC model are also listed in Table~\ref{tab:par}.

\begin{table*}
\scriptsize
\begin{center}
\caption{Values of flux data points of SEDs}
\begin{tabular}{cccccccccccccc}
\hline\hline
& \multicolumn{13}{c}{Whole data} \\
Flux& 13.90(14) & 12.98(10) & 12.25(8) & 10.59(8) & 7.98(8) & 5.50(9) & 3.22(9) & 1.46(9) & 0.71(8) & 0.46(9) & 0.27(10) & 0.56(6) & 0.13(2) \\
TS  & 31232.1 & 39091.2 & 48564.4 & 47672.0 & 33271.5 & 17115.8 & 6326.9 & 1574.0 & 420.3 & 141.5 & 36.4 & 58.7 & 6.8 \\
\hline
& \multicolumn{13}{c}{Dip data} \\
Flux& 12.31(44) & 11.76(26) & 11.76(21) & 10.21(21) & 7.74(21) & 5.39(23) & 3.13(24) & 1.69(50) & 0.55(20) & 0.70(29) & 0.28(27) & 0.72(19) & 1.06$^{\rm a}$ \\
TS  & 3777.0 & 4975.9 & 6941.4 & 6806.6 & 4856.0 & 2455.7 & 959.7 & 289.0 & 35.5 & 39.7 & 6.7 & 9.3 & --- \\
\hline
\end{tabular}
\label{tab:sed-data}
\end{center}
{\bf Notes. }{Values of flux data points of SEDs for the whole and dip data in 0.1--500.0~GeV
              based on LP model, the fluxes in units of $10^{-11}$~erg~cm$^{-2}$~s$^{-1}$.
              Last data point$^{\rm a}$ of dip data is the 95\% flux upper limits.
              Numbers in parentheses represent uncertainties on the last digit.}
\end{table*}

Based on the new model, we performed spectral analysis to derive the SED of \ls~in 0.1--500.0~GeV.
The whole LAT data was segmented into 13 equally logarithmically spaced energy bins. The data points
of SED were extracted by performed the maximum likelihood analysis.
\ls's SED is shown in Figure~\ref{fig:sed} and the accurate numbers are listed in Table~\ref{tab:sed-data}.
For ease of comparison, we show the LP and PLEC models in Figure~\ref{fig:sed} with
black solid and blue dotted lines.
From them, we know that the LP model performs better than PLEC for target's \gr~emission.

\begin{figure}
\centering
\includegraphics[angle=0,scale=0.67]{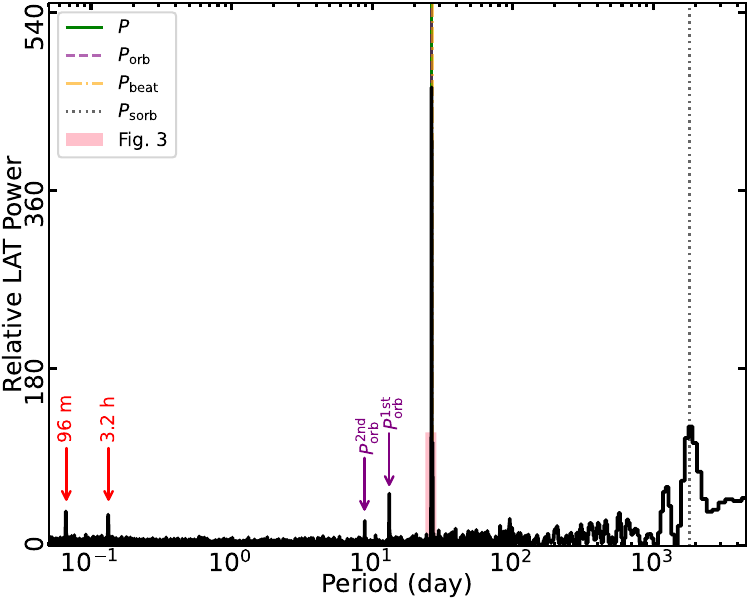}
\caption{LSP power spectrum (black histogram) constructed from 0.1--500.0 GeV AP light curve of \ls.
         The three periodicities of $P_{\rm new}$, $P_{\rm orb}$, and $P_{\rm beat}$ are too close,
         resulting in overlap, for clarity, please refer to Figure~\ref{fig:Po}.
         The $P_{\rm sorb}$ is marked with a gray dotted line.
         The orbital and survey repeat periods are marked with two red arrows,
         and the first and second harmonics of $P_{\rm orb}$ are shown with two purple arrows.
         The pink shaded region stands for the portion of power spectrum shown in Figure~\ref{fig:Po}.}
\label{fig:all-lsp}
\end{figure}

In order to reveal the \gr~emissions around \ls, a TS map with a region of $4^{\circ}\times4^{\circ}$
was created by employed \emph{gttsmap} based on the new model file.
And a residual TS map was also created to exclude the
contamination from the new possible nearby \gr~sources, that not included in 4FGL-DR4.
We show the two TS maps in Figure~\ref{fig:tsmap}.
From them, we believe that the \gr~events around \ls~are well described by the new model,
and no new \gr~source has been identified beyond 4FGL-DR4.

\subsection{Timing Analysis}
\label{sec:timing}
An initial light curve was constructed by employed a modified version of aperture photometry (AP) method
centered at 4FGL J0240.5+6113. Taking into account the instrument performance of LAT
to maximize the signal-to-noise ratio, an aperture radius of $3^\circ.16$ is adopted with
a selection criteron of an angle $\theta<$ max(6.68$-$1.76log$_{10}(E_{\rm MeV}), 1.3)^{\circ}$,
as that performed in \citet{abd+10}.
The light curve has an energy range of 0.1--500.0~GeV with time bins of 600~s. We excluded the periods when
4FGL J0240.5+6113 was within $5^\circ$ of the Sun and Moon by \emph{gtmktime}. Exposures were calculated using
\emph{gtexposure} to mitigate the impact of significant exposure variations across different time bins. 
And \gr s arrival times are also barycenter corrected using \emph{gtbary}. We assigned weights to events
with their probabilities of originating from 4FGL J0240.5+6113 by employed \emph{gtsrcprob} based on the new model.
The light curve was then constructed by summing these probabilities, as opposed to simply counting
the number of photons within the aperture in each time bin \citep{k11,J1018+12,cor+19}.

Power spectrum was created for the AP light curve
by employed a method of a Lomb–Scargle periodogram \citep[LSP;][]{l76,s82},
and we show it in Figure~\ref{fig:all-lsp}. It covers a frequency range from $f_{\rm max}$~=~1/0.05
day$^{-1}$ to the entire \emph{Fermi}-LAT observations ($f_{\rm min}$~=~1/5525 day$^{-1}$),
and the number of independent frequencies (i.e., the trial factor) was calculated by
$N=(f_{\rm max}-f_{\rm min})/\delta f$ = 110499, where $\delta f$ is frequency resolution determined
by the length of the \emph{Fermi}-LAT observations.
In the spectrum, the \gr~periodicities of \ls~that have been reported in previous works,
i.~e. the orbital ($P_{\rm orb}$), superorbital ($P_{\rm sorb}$),
and orbital–superorbital beat periods ($P_{\rm beat}$), reveal at $26.493\pm0.058$, $1817.54\pm252.84$,
and $26.927\pm0.055$~day, respectively. We marked them with purple dashed, gray dotted,
and orange dashed-dotted lines, respectively.
Interestingly, besides these \gr~periodicities, an additional periodic signal also reveals
at $26.301\pm0.037$~day, this new periodic signal ($P_{\rm new}$) is marked with a green solid line
in Figure~\ref{fig:all-lsp}.
Their uncertainties of the \gr~periodicities were taken from their half-widths at half-maximum of
each power peak.
In our AP analysis the background \gr s are not modeled out for each time bin, hence two artifact signals
(marked with red arrows) originate from the 96 minutes orbital period and the survey repeat period at
twice of them of \emph{Fermi}
satellite\footnote{https://fermi.gsfc.nasa.gov/ssc/data/analysis/LAT\_caveats\_temporal.html}
can be seen at their corresponding periods.

\begin{figure}
\centering
\includegraphics[angle=0,scale=0.68]{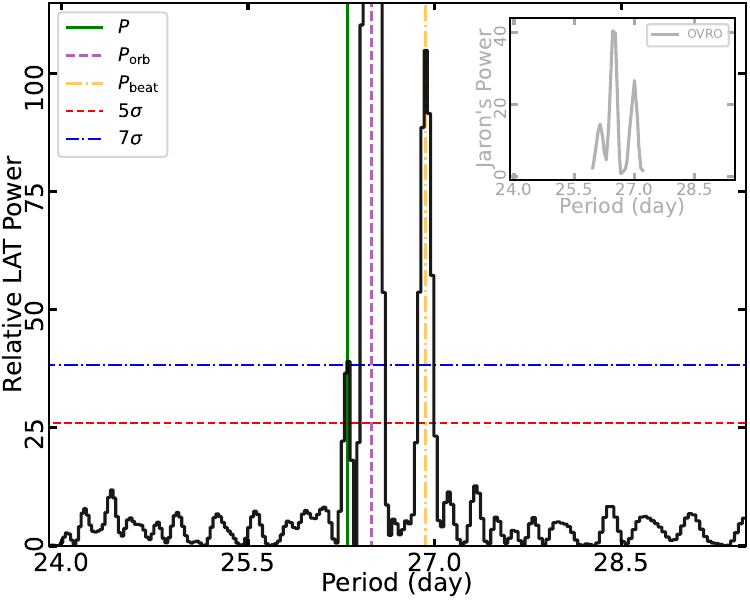}
\caption{LSP power spectrum zoomed in Figure~\ref{fig:all-lsp} (i.e. the pink shaded region).
         The new detected periodicity $P_{\rm new}$ is marked with a green solid line,
         and others for $P_{\rm orb}$ and $P_{\rm beat}$ are shown with purple dashed and orange
         dashed-dotted lines, respectively. The red dashed and blue dashed-dotted lines stand for
         $5\sigma$ and $7\sigma$ confidence levels.
         The inset plot shows that the schematic LSP power of \ls~drawn from
         the Figure~3~(d) of \citet{jar+18}.}
\label{fig:Po}
\end{figure}

Because the three periodicities of $P_{\rm new}$, $P_{\rm orb}$, and $P_{\rm beat}$ have similar periods and
result them in overlap each other in Figure~\ref{fig:all-lsp},
for clarity we zoom in the power spectrum in the pink shaded region in Figure~\ref{fig:Po}.
The heights of the peaks of $P_{\rm new}$, $P_{\rm orb}$, $P_{\rm sorb}$ and $P_{\rm beat}$
are $\sim39.1$, $\sim465.9$, $\sim121.5$ and $\sim105.0$ compared to the mean power level, respectively.
The normalization method utilized here is detailed in \citet{hb86}.
Using the method provided by \citet{l76} and \citet{s82}, the probability ($p_{_{\rm lsp}}$) to obtain
the power level of $P_{\rm new}$ equal or higher than $39.1$ from a chance fluctuation (a noise)
is $\sim1.0\times10^{-17}$. This method is often used for detecting periodic signals in the white noise,
as demonstrated in the search for \gr~binary systems by \citet{cor+16,cor+19,lat+12}.
Taking into account the trial number $N$, the False Alarm Probability (FAP) is estimated
at FAP=$1-(1-p)^N\sim p\times N\sim1.1\times10^{-12}$, which corresponding to
a $\sim7.1\sigma$ confidence level. In Figure~\ref{fig:Po}, we show $5\sigma$ and
$7\sigma$ confidence levels with red dashed and blue dashed-dotted lines, respectively.
Interestingly, in radio OVRO data, a potential signal exhibits a period ($26.16\pm0.11$~day,
the error derived from the half-widths at half-maximum of the power peak) similar to the new signal.
For easy reference, we show the schematic LSP power of \ls~in the inset
plot of Figure~\ref{fig:Po}, that drawn from \citet{jar+18}.

In addition, a potential periodic signal of nearby \gr~sources can also cause a modulation for
\ls~because of the broad PSF of the \emph{Fermi}-LAT, particularly at lower energies.
To prevent this situation, we also constructed the power spectra for the two closest sources
(4FGL~J0248.4+6021 and 4FGL~J0243.3+6319) with the same process, and no similar signal was
identified for the new periodic signal claimed here.

\begin{figure}
\centering
\includegraphics[angle=0,scale=0.68]{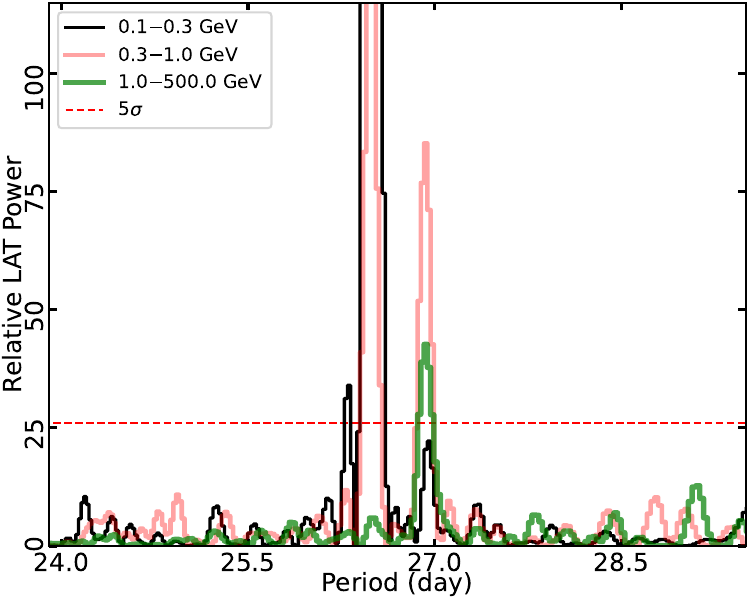}
\caption{\ls's energy-dependent LSP power spectra in 0.1--0.3, 0.3--1.0, and 1.0-500.0~GeV.
         The black, red, and green histograms stand for the LSP power spectra constructed
         with 0.1--0.3, 0.3--1.0, and 1.0--500.0~GeV AP light curves, respectively.}
\label{fig:ende}
\end{figure}

To explore the energy-dependent periodicity of $P_{\rm new}$, we also created three power spectra
based on their AP light-curves in 0.1--0.3, 0.3--1.0, and 1.0--500.0~GeV.
And they are shown in Figure~\ref{fig:ende} with black, red, and green histograms, respectively.
From it, we know that their power peak of $P_{\rm new}$ signal gradually decreases as the
energy increases across the three energy ranges.

\begin{figure}
\centering
\includegraphics[angle=0,scale=0.93]{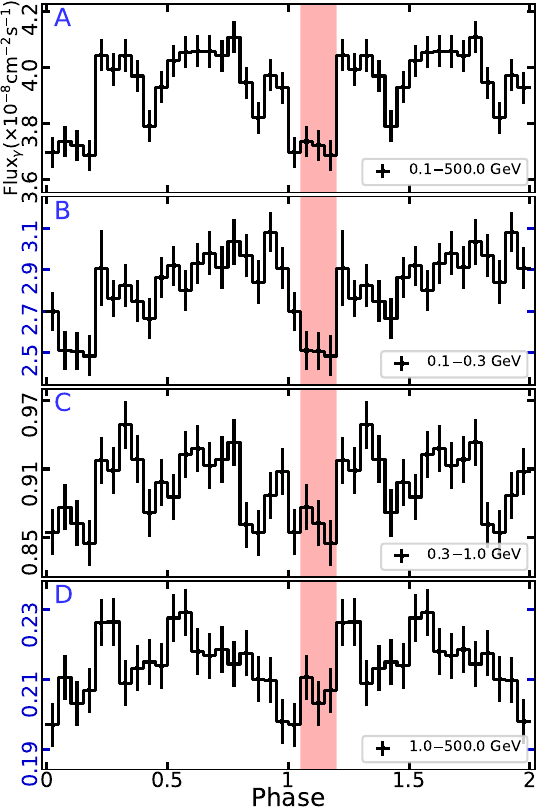}
\caption{Phase-resolved light-curves. Flux varies with the corresponding phase
         in the four energy intervals in 0.1--500.0, 0.1--0.3, 0.3--1.0, and 1.0--500.0~GeV.
         And four panels share same y-axis labels. For clarity two period cycles are shown.
         The pink shade denotes the dip in $\phi=0.05-0.20$.}
\label{fig:phlc}
\end{figure}

\subsection{Phase-resolved Analysis}
\label{sec:phlc}
On the basis of period of $P_{\rm new}$, we divided the 0.1--500.0~GeV \emph{Fermi}-LAT events into 20 phase
intervals and performed likelihood analysis for each bin to obtain the phase-resolved light-curve
based on the new model file, except only freeing the flux normalizations for the sources
within 5$^{\circ}$ and two diffuse components.
The folded light-curve is shown in Figure~\ref{fig:phlc} (A) with a phase zero corresponding to
MJD~43,366.275, as it in \citet{g02,cmn+23}.
From it, an obvious dip in $\phi=0.05-0.20$ can be seen in the folded light-curve.

In order to investigate the phase-resolved light-curve depending on energy,
we constructed phase-resolved light-curves for the three previous energy intervals. And we show them
in B-D panels respectively. Their amplitudes of the variabilities,
i.e. the maximum flux minus the minimum, for the four phase-resolved light-curve are
0.42, 0.60, 0.10, and 0.03$\times10^{-8}$~photons~cm$^{-2}$~s$^{-1}$.
As shown in the Figure~\ref{fig:phlc}, the amplitudes of the phase-resolved
light-curves of $P_{\rm new}$ in the three energy ranges decrease as the energy increases.

We also carried out a maximum likelihood analysis for the events in the phase intervals
in the dip (pink shade in Figure~\ref{fig:phlc}).
Its best-fit LP model is shown in Figure~\ref{fig:sed} with a red dashed line
and the parameters are listed in Table~\ref{tab:par}.
A spectral analysis was performed for the data in the dip, the data points of SED are
displayed in Figure~\ref{fig:sed} and colored red. For the SED of dip data, we retained data points with TS values
$\geqslant$5, while others are displayed their 95\% flux upper limits.
Based on Figure~\ref{fig:sed} and Table~\ref{tab:sed-data}, it is evident that there are no significant differences in the SEDs
between the entire LAT data and the dip data.

\section{Summary and Discussion}
\label{sec:dis}
\ls~is one of the unusual \gr~binaries and composed of a massive star and a rapidly rotating neutron
star \citep{wen+22}. The main of the electromagnetic emissions from \ls~are at MeV-GeV
energies \citep{d15,dub+17}.
The principal emission mechanisms of this system are thought to be \gr~emissions that could
originate from interactions between the relativistic wind coming from a rapidly rotating neutron
star \citep{d06} and the stellar wind from its companion, or from the relativistic jets generated
by accretion onto a neutron star or black hole \citep{mr98}.
In \gr s, for \ls, three periodicities corresponding to the orbital, superorbital,
and orbital–superorbital beat periods have been extensively discussed in previous works.
Using the events from \emph{Fermi}-LAT spanning from 2008 August to 2023 September,
we carried out a timing analysis in 0.1--500.0~GeV for \ls, and a new periodic
signal with a period of $26.301\pm0.037$~day was detected at a $\sim7\sigma$ confidence level.
Interestingly, in radio OVRO data, a potential signal (not significant) with a similar period
is revealed at $26.16\pm0.11$~day \citep{jar+18}. Furthermore, their error ranges also partially overlap.
Their results further strengthen our \gr~findings independently.
The new signal is relatively weak compared to the other known periodicities.

As reported by \citet{cmn+23}, two periods of \ls, corresponding to the orbital and
beat orbital/superorbital periods, exhibit strong energy dependence.
We also conducted an energy-dependent analysis for the new signal.
Our results in the three same energy intervals indicates that 
the power spectrum peak of the new signal gradually decreases from the low energy range
to high. We speculate that the cause of our results may be due to statistical effects,
as there are significantly more photons at lower energies compared to higher energies.

From the phase-resolved light curves of energy dependent intervals (B, C, and
D panles of Figure~\ref{fig:phlc}), we know that their amplitudes of the light-curve
decreases from low energy to high. From them, an obvious dip
can be seen at $\phi=0.05-0.20$, especially in the whole and low-energy data (A and B
panels of Figure~\ref{fig:phlc}).
While the orbital periodic periodicity of \ls~is characterized by a single peak in radio to
X-ray and $\gamma$-ray bands \citep{cmn+23,xwt17}.
Hence, we speculate that the origin of the new periodic signal may be different
from the three periodicities discussed in previous literature.
Perhaps long-term, high-cadence, and multi-wavelength observations can reveal it
in the light curve.

\begin{figure}
\centering
\includegraphics[angle=0,scale=0.66]{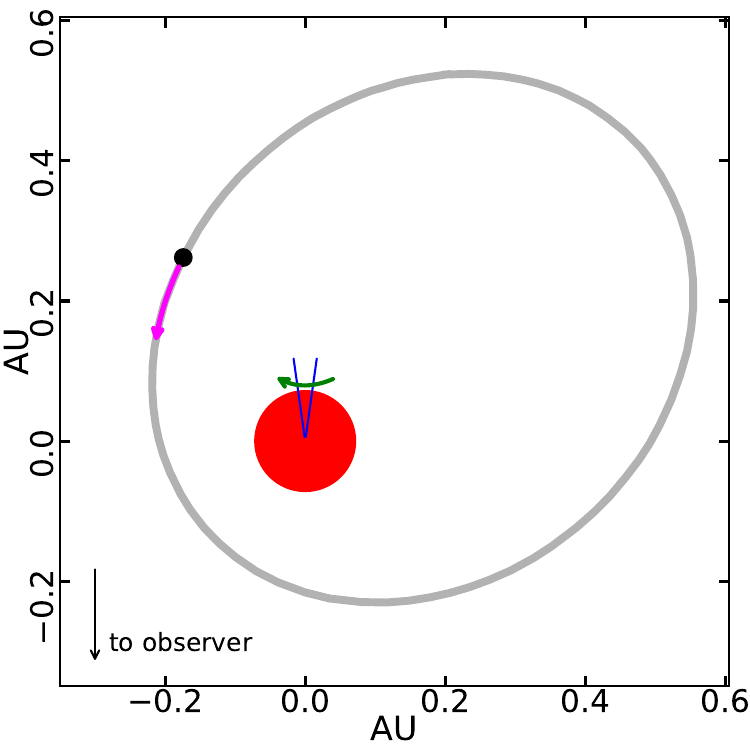}
\caption{Schematic illustration of the orbit of \ls, drawn from \citet{d13}.
         The purple arrow indicates the direction of the neutron star's motion,
         the blue lines represent the spin axis of the Be star, and the green arrow
         denotes the direction of the spin axis's precession.}
\label{fig:orb}
\end{figure}

In our analysis, we checked the various combinations of frequencies involving known periods
(i.e. $P_{\rm average}=\frac{2}{f_{\rm x}+f_{\rm y}}$, $P_{\rm beat}=\frac{1}{f_{\rm x}-f_{\rm y}}$)
or possible harmonics.
We found that the new periodic signal is not a combination of frequencies from any known periods.
It appears to be a new, independent periodic signal.
The spin precession of Be stars in X-ray binaries has long been studied \citep{lo00,bbc+00,od01,mpt+11}.
We suggest that the Be star in the \ls~has a spin axis's precession,
and the precession direction is opposite to the orbital motion of the neutron star.
To facilitate understanding of our hypothesis,
we show this scenario and the Schematic illustration of orbit of \ls~in Figure~\ref{fig:orb}.
Then we can derive the $P_{\rm prec}$ with a formula of $1/P_{\rm new}=1/P_{\rm orb} + 1/P_{\rm prec}$.
The period of $P_{\rm prec}$ is derived to be $\sim3629$~day, which falls within the timescale of
Be star's precession estimated by \citet{m23}.
Therefore we speculate that the newly discovered periodic signal ($P_{\rm new}$) may originate
from a coupling effect between the orbital period ($P_{\rm orb}$) and the retrograde stellar
precession period ($P_{\rm prec}$).

Regardless, the real originating for the new periodic signal still remains unclear.
Based on our results, \ls~may possess some properties that are currently unknown.
More observations in the multi-wavelengths are encouraged to reveal the origin of the new periodic signal
reported here.

\begin{acknowledgments}
We thank anonymous referee for very helpful suggestions and
Z. Wang for discussion about potential origins of the new periodic signal.
This work is supported in part by the National Natural Science Foundation of 
China Nos.~12163006 and 12233006, the Basic Research Program of Yunnan Province 
No.~202201AT070137, and the joint foundation of Department of Science and Technology
of Yunnan Province and Yunnan University No.~202201BF070001-020.
P.F.Z. acknowledges the support by the Xingdian Talent Support Plan - Youth Project. 
\end{acknowledgments}

\bibliographystyle{aasjournal}
\bibliography{aas}
\end{document}